# Single Diamond Structured Titania Scaffold

Chao Wang[1], Congcong Cui[1], Quanzheng Deng[1], Chong Zhang[1], Shunsuke Asahina[2], Yuanyuan Cao[3], Yiyong Mai[4], Shunai Che[1,4], Lu Han[1]*

[1] School of Chemical Science and Engineering, Tongji University, 1239 Siping Road, Shanghai, 200092, China.
[2] SMBU, JEOL, Akishima, Tokyo 196-8558, Japan.
[3] School of Materials Science and Engineering, East China University of Science and Technology; Shanghai, 200237, China.
[4] School of Chemistry and Chemical Engineering, State Key Laboratory of Composite Materials, Shanghai Key Laboratory for Molecular Engineering of Chiral Drugs, Shanghai Jiao Tong University, 800 Dongchuan Road, Shanghai, 200240, China.
*e-mail: luhan@tongji.edu.cn

**Abstract:** The single diamond (SD) network, discovered in beetle and weevil skeletons, is the "holy grail" of photonic materials with the widest complete bandgap known to date. However, the thermodynamic instability of SD has made its self-assembly long been a formidable challenge. By imitating the simultaneous co-folding process of nonequilibrium skeleton formation in natural organisms, we devised an unprecedented bottom-up approach to fabricate SD networks via the synergistic self-assembly of diblock copolymer and inorganic precursors and successfully obtained tetrahedral connected polycrystalline anatase SD frameworks. A photonic bandstructure calculation showed that the resulting SD structure has a wide and complete photonic bandgap. This work provides an ingenious design solution to the complex synthetic puzzle and offers new opportunities for biorelevant materials, next-generation optical devices, etc.

The single diamond (SD) network (*dia*-net), resembling the cubic diamond lattice with a network enclosed by continuous saddle-shaped surface, has been long sought for photonic materials[1-3]. This structure possesses the widest complete photonic bandgap reported so far, and a complete photonic bandgap can be obtained using known materials at a refractive index contrast of ~1.9[4]. The superior optical properties of SD make it stand apart from other palette of natural or artificial photonic structures ever discovered[5,6], enabling SD sparkling in various applications such as photonic crystals[7,8], light-harvesting applications[9,10], optical waveguides[11,12], laser resonators[13], etc. Although SD structures have been shown to be a versatile sources of biomorphic scaffold designs[14,15], their formation mechanism remains elusive and beyond current synthetic conceptions[16,17]. Only the thermodynamically favoured double diamond (DD) structures, either the triply periodic diamond minimal surface or the corresponding interpenetrated-DD skeletons, generally appear in artificial syntheses, such as phase-separated block copolymeric aggregates and amphiphilic lipidic micelles[18-21]. Unfortunately, these DD structures have been shown to abolish the photonic properties due to the changes in symmetry[4].

To realize the chemical synthesis of SD, it is indispensable to go beyond the limitation of thermodynamic principles. Although the SD can be formed by replicating biological networks[22] or initially formed diamond minimal surface by sealing one of the double networks as hard template[23], there are tedious steps required, and the structure is greatly limited by the template (i.e., randomly oriented domains in biological template sources). One general thought to solve this synthesis challenge is to introduce specific building blocks with designed molecular geometry constraints, which has been reported in thermotropic liquid crystal[24] and colloidal superlattice[25]. However, the photonic properties of liquid crystals are greatly limited by the low dielectric contrast and the small unit cell parameter; and the spherical-based structure of colloids is deviated from the ideal continuous dielectric property of the biological SD surface. Another approach is to form an alternating version of SD, and it is believed that ABC triblock terpolymers must be employed to eliminate the inversion symmetry of the two networks in DD[26,27], while the complex syntheses of the triblock terpolymers and the tedious block-selective etching and backfilling processes needed would be major drawbacks[28]. Therefore, there is currently no artificial method that can efficiently prepare SD through a one-step self-assembly as what happens in natural organisms[29], and the synthesis of SD with a complete photonic bandgap has not yet been realized.

Inspired by the remarkable simultaneous co-folding formation process of the nonequilibrium state network structures in natural species, such as the single gyroid networks[29], we design a novel, versatile and efficient bottom-up approach for fabricating SD structured titania network. To achieve a biologically similar co-folding process involving cooperative organization of endoplasmic reticulum template and the selectively deposition of chitin as in butterfly wing scales, we propose a synthetic strategy that employ dual control of thermodynamics and kinetics. An equilibrium diamond minimal surface is formed by simultaneous co-folding like organization of diblock copolymer soft templates and titania precursors, and a biologically-resembling kinetically controlled nucleation and crystal growth of titania species specifically within one of the double networks, enabling the formation of a single network.

Fig. 1 illustrates our synthetic strategy. The synthesis employed the commonly used diblock copolymer poly(ethylene oxide)-*b*-polystyrene ($PEO_{45}$-*b*-$PS_{241}$) as a soft template and titanium diisopropoxide bis(acetylacetonate) (TIA) as an inorganic source in a mixture of tetrahydrofuran (THF) and HCl aqueous solution. The $PEO_{45}$-*b*-$PS_{241}$ was synthesized by atom-transfer radical polymerization and exhibited a total molar mass of 27.1 kg/mol (*Đ* = 1.12) and volume fractions of $f_{PEO}$ = 6.9% and $f_{PS}$ = 93.1% (Supplementary Information Figs. S1-S4 and Table S1). The Flory-Huggins interaction parameter $\chi N \approx 43.1$ suggested strong microphase separation (Supplementary Information Tables S2 and S3).

The template macromolecules were initially dissolved in the common solvent THF due to the similar solubility parameters[30] of PS ($\delta$ = 19.3 $[MPa]^{1/2}$), PEO ($\delta$ = 21.2 $[MPa]^{1/2}$) and THF ($\delta$ = 18.6 $[MPa]^{1/2}$) (Fig. 1a, Supplementary Information Table S2). After addition of the $H_2O$/HCl ($\delta_{water}$ = 47.80 $[MPa]^{1/2}$), the hydrophobic PS segments were compatible with the THF-rich phase, while water was associated with the hydrophilic PEO region. After evaporation of the solvents, the diamond minimal surface was formed by the block copolymer, where the PS segment formed the interface, separating the space in halves and resulting in two sets of PEO skeletons (channels). Simultaneously, the titania species were hydrolysed and condensed to form hydrophilic oxo oligomers $TiX_x(OH)_y$-$O_{2-(x+y)/2}$ (X = OR or Cl) and interacted with the polymer through hydrogen bonding with the oxygen atoms in the PEO chains[31] (Fig. 1b).

Initially, the inorganic species were distributed in both sets of PEO skeletons. However, the reaction kinetics can be regulated with $H_2O$ and HCl[32]. Upon addition of a small amount of $H_2O$, slow hydrolysis and condensation of TIA were expected, thus reducing the number of nucleation sites (Fig. 1c). Although these nucleation sites were randomly located in both sets of skeleton channels, the small number made them sparsely distributed in the polymer matrix, where they served as the core for enrichment and growth of the titania precursors and formed the independent SD domains (Fig. 1d). As the amount of H$_2$O was increased, there were many TiO$_2$ nucleation sites formed in both skeletons, resulting in the formation of DD domains with two identical continuous TiO$_2$ networks (Fig. 1f-1g). Finally, the polymer template was removed by calcination, and the amorphous titania framework was transformed to a crystalline form (Fig. 1e and 1h).

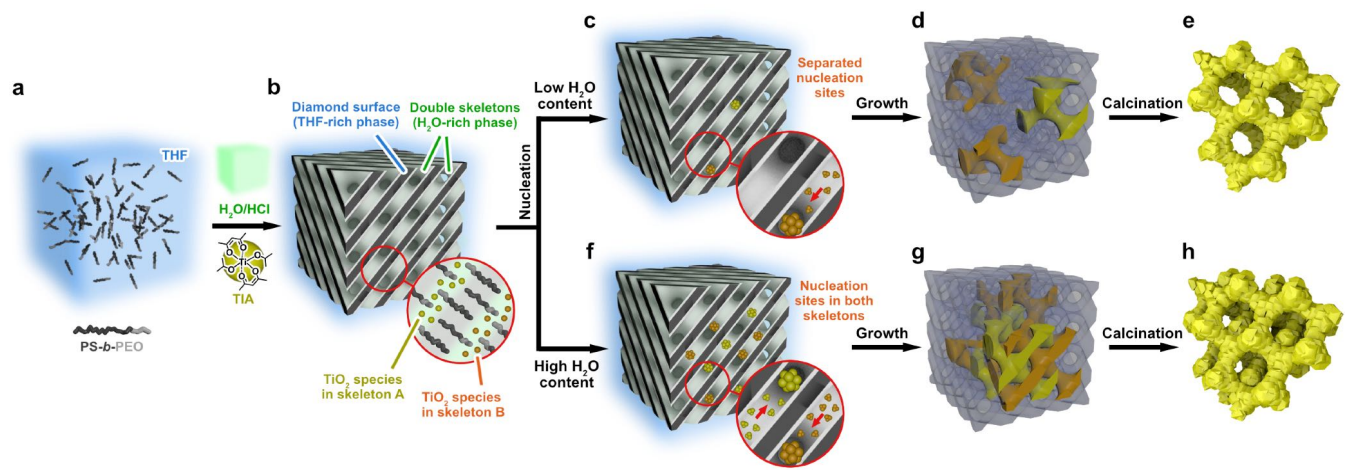

**Fig. 1. Schematic drawing of the synthetic strategy for the SD and DD titania scaffolds. a**, A uniform PEO$_{45}$-*b*-PS$_{241}$ solution was formed in THF. **b**, With the addition of H$_2$O/HCl and the titania precursor TIA, the diamond minimal surface matrix was formed by the block copolymer with solvent evaporation, where the PS segment formed the interface dividing the space into two interpenetrating skeletons associated with the PEO regions. The titania species were simultaneously hydrolysed and formed hydrophilic oxo oligomers in the PEO chain region. **c**, With the addition of a small amount of H$_2$O, a limited number of nucleation sites are formed, resulting in separation of the nucleation sites. **d**, The enrichment and growth of titania oligomers led to the formation of independent SD domains. **e**, After calcination, the polymer template was removed, leaving the SD anatase titania scaffolds. **f**, When the amount of H$_2$O was increased, a large number of titania nucleation sites existed in both double skeletons. **g**, The DD titania network formed embedded in the polymer matrix. **h**, The DD anatase titania scaffolds were formed after calcination.

Based on this scheme, a series of samples was produced by changing the HCl (3 M) content (Supplementary Information Fig. S5). No ordered structure was obtained when the mass ratio of HCl (3 M)/THF was 0.05 (Supplementary Information Fig. S6). The TiO$_2$ scaffold with an SD structure was formed when the HCl (3 M)/THF mass ratio ranged from 0.075 to 0.125. The DD structure was formed with HCl (3 M)/THF mass ratios of 0.15 to 0.2, which finally converted into a cylindrical phase at much higher HCl (3 M)/THF mass ratios from 0.25 to 0.375.

The typical SD sample synthesized with the HCl (3 M)/THF mass ratio of 0.1 was characterized by a series of detailed analyses. The small-angle X-ray scattering (SAXS) profile (Fig. 2a) of the as-synthesized SD sample exhibited several well-resolved reflections within the $q$ range of 0.1-1.0 nm$^{-1}$, representing a highly ordered structure. Notably, both the polymer matrix with a diamond minimal surface and the embedded SD TiO$_2$ network contributed to these reflections, as confirmed by electron microscopy (*vide post*). All of the peaks except the first peak (blue tick markers in Fig. 2a) were assigned to the 110, 111, 200, 211… reflections of cubic space group $Pn\bar{3}m$ (No. 224) with the unit cell parameter $a \approx 45$ nm. Moreover, the embedded SD TiO$_2$ skeleton was recognized by the intense 111 reflection centred at 0.12 nm$^{-1}$. Due to the geometrical relationship of the SD and the diamond minimal surface, other SD reflections were indexed to $2h2k2l$ of the polymer matrix due to the doubled unit cell size of $a \approx 90$ nm (Supplementary Information Fig. S7). After calcination, the polymer template was removed, and the amorphous TiO$_2$ was transformed into the anatase crystalline form. Therefore, the diffraction peaks moved to higher $q$ values due to shrinkage of the TiO$_2$, and the reflections were assigned to the space group $Fd\bar{3}m$ with a lattice parameter of $a \approx 60$ nm.

The crystallinities of the samples at the atomic scale were revealed by the wide-angle X-ray diffraction (XRD) patterns (Fig. 2b). The as-synthesized SD sample showed a broad peak at approximately $2\theta = 20°$, corresponding to the amorphous structure, while some weak reflections suggested a small proportion of the crystalline state. After calcination, several intense peaks appeared, which were consistent with the anatase phase (JCPDS File No. 21-1272). The average size of the crystalline TiO$_2$ nanocrystals was estimated to be 15-20 nm.

The samples were also investigated by scanning electron microscopy (SEM). The SEM image of the as-synthesized sample suggested a single crystal domain with a highly regular structure (Fig. 2c). The high magnification image (Fig. 2d) indicated the typical {111} plane of the diamond minimal surface with several growth steps. The SEM images taken from other orientations were also consistent with the diamond surface (Supplementary Information Fig. S8). To determine the three-dimensional (3D) organization of the polymer matrix, transmission electron microscopy (TEM) images were taken along different zone axes from the thin edges of the as-synthesized sample. The reconstructed 3D electrostatic potential distribution clearly revealed a typical diamond minimal surface-type structure (Supplementary Information Fig. S9 and Table S4).

To verify the presence of the SD skeleton embedded in the polymer matrix, the inner structural features of the as-synthesized particle were determined by cross-sectional polishing with accelerated argon ion beam milling (inset of Fig. 2e). Since the heavy atoms backscattered the electrons much more efficiently than the light elements, backscattered electron (BSE) imaging was performed to enhance the contrast of the TiO$_2$ region. As shown in Fig. 2e, the particle size exceeded 100 μm across, and there are numerous bright regions uniformly distributed throughout the whole structure, corresponding to the TiO$_2$ networks embedded in the polymer matrix. In the high-magnification BSE image (Fig. 2f), these TiO$_2$ domains showed a clear morphology with distinct crystal facets, and energy dispersive X-ray spectroscopy (EDS) mapping (Fig. 2g) indicated brighter contrast consistent with the TiO$_2$ region. The integration of the TiO$_2$ species in only one of the two skeletons in the matrix was revealed by high-magnification secondary electron imaging (Fig. 2h) and BSE imaging (Fig. 2i), from which every other channel was filled with TiO$_2$; the remaining half remained unoccupied, suggesting the formation of an SD TiO$_2$ network.

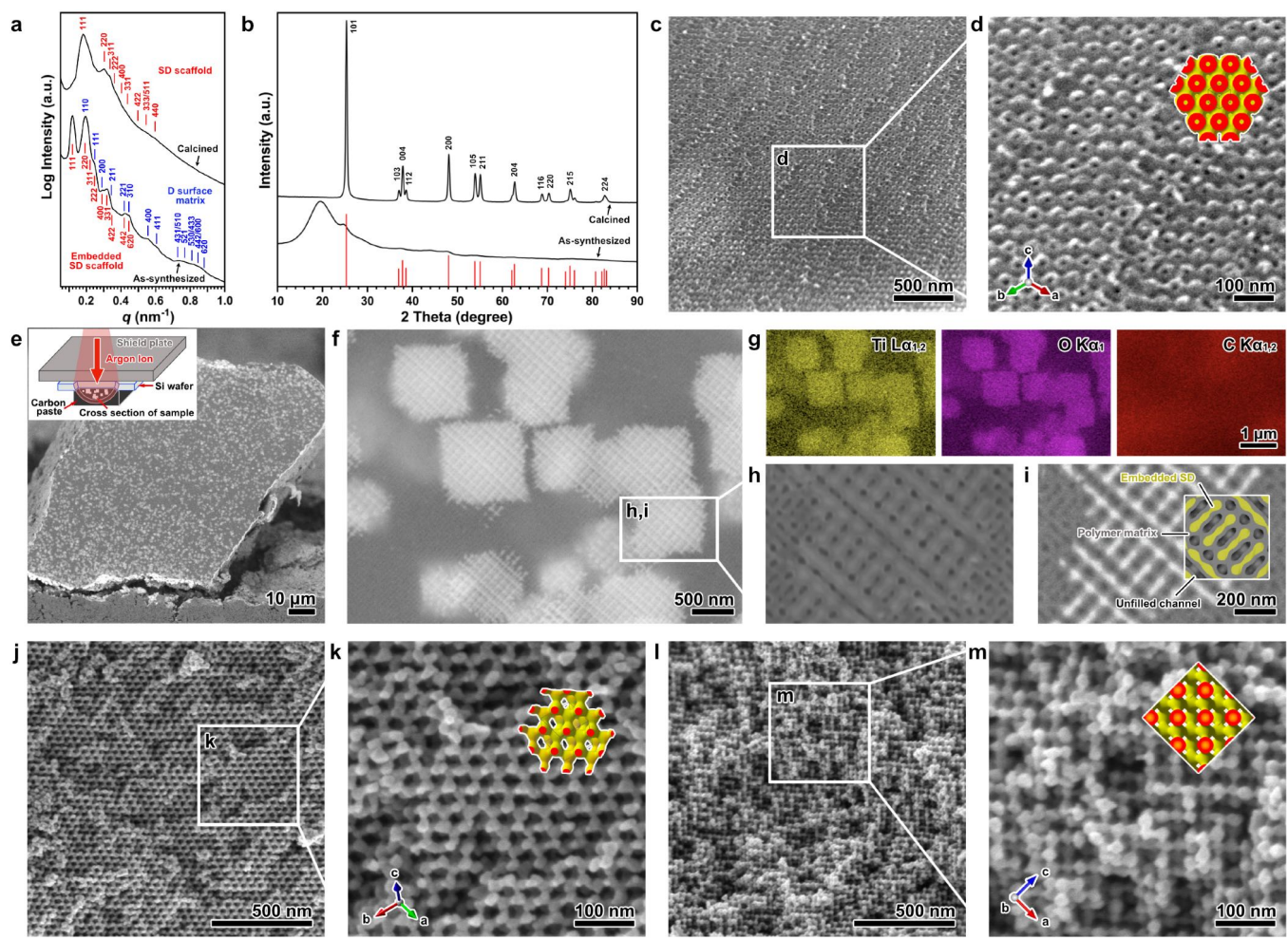

**Fig. 2. Structural characterizations of the DD and SD structures. a**, SAXS patterns of the as-synthesized and calcined SD samples. **b**, Wide-angle XRD patterns of the as-synthesized and calcined SD samples. **c, d**, Low- and high-magnification SEM images of the as-synthesized sample taken from the {111} lattice plane. **e**, Low-magnification cross-sectional BSE image of the as-synthesized sample. The inset shows a schematic drawing of the cross-section polishing process performed with argon ion beam milling to create artefact-free cross sections. The bright contrast corresponds to the $TiO_2$ region. **f**, High magnification BSE image of the cross-section. **g**, EDS mapping of **f**. **h, i**, High-magnification secondary electron image and BSE image of the cross section shown in the white rectangle in **f**. **j, k**, Low- and high-magnification SEM images of the calcined SD taken from the {111} plane. **l, m**, Low- and high-magnification SEM images of the calcined SD scaffold observed from the {010} plane.

After calcination, the block copolymer matrix was removed, and the internal SD $TiO_2$ networks were revealed. The size of the SD structure with a single crystal domain in the calcined sample exceeded 10 μm, indicating long-distance order (Supplementary Information Fig. S10). The high magnification SEM images (Fig. 2j-2m) revealed single network scaffold structures with fourfold connectivity at the nodal sites, a typical characteristic of the SD structure. The SEM images taken from other directions are shown in Supplementary Information Fig. S11.

The 3D structure and fine details of the calcined SD sample were then investigated by TEM. The low magnification TEM images showed that the SD lattice was highly ordered over several microns without the presence of double network structures (Supplementary Information Fig. S12). The TEM images and the corresponding Fourier diffractograms (FDs) taken from the [010], [110], [111] and [112] directions of the calcined SD sample are shown in Fig. 3a-3d. The extinction conditions from the FDs can be summarized as $\{hkl: h + k, h + l, k + l = 2n\}$, $\{0kl: k + l = 4n, k, l = 2n\}$, $\{hhl: h + l = 2n\}$, and $\{00l: l = 4n\}$, indicating two possible space groups, $Fd\bar{3}$ (No. 203) and $Fd\bar{3}m$ (No. 227). The unit cell parameter calculated from the FDs ($a \approx 60$ nm) was consistent with the SAXS and SEM results. Then, the electrostatic potential map was calculated by electron crystallographic 3D reconstruction to reveal the configuration of the SD structure. Both amplitudes and phases were extracted from the FDs with the crystallographic image-processing software CRISP[33], and the plane groups $p4mm$, $c2mm$, $p6mm$ and $p2mm$ were determined from the representative zone axes. Therefore, the space group $Fd\bar{3}m$ was uniquely chosen according to the plane group symmetry. Each 2D projection was adjusted to the common crystal origin at the inversion centre (origin choice 2 of the space group). The crystal structure factors were then merged into one dataset by normalizing upon scaling the common reflections. The 3D electrostatic potential map $\varphi(x,y,z)$ was calculated after correcting the contrast transfer function with a Wiener filter to avoid zero division (Supplementary Information Table S5). The threshold for the equi-electrostatic potential surface was directly determined from the TEM images. Fig. 3f and 3g present the reconstructed 3D map of the unit cell. The stick model superimposed on the reconstructed volume clearly showed the tetrahedrally linked SD networks. The slice view of the (111) facets (Fig. 3h) was also in good agreement with the SEM observations (Fig. 2j and 2k). To verify the structure, the TEM images were simulated with a 3D continuum model[34] employing nodal approximation[35] (insets of Fig. 3a-3d). The good agreement suggested the faithfulness of the structural solution.

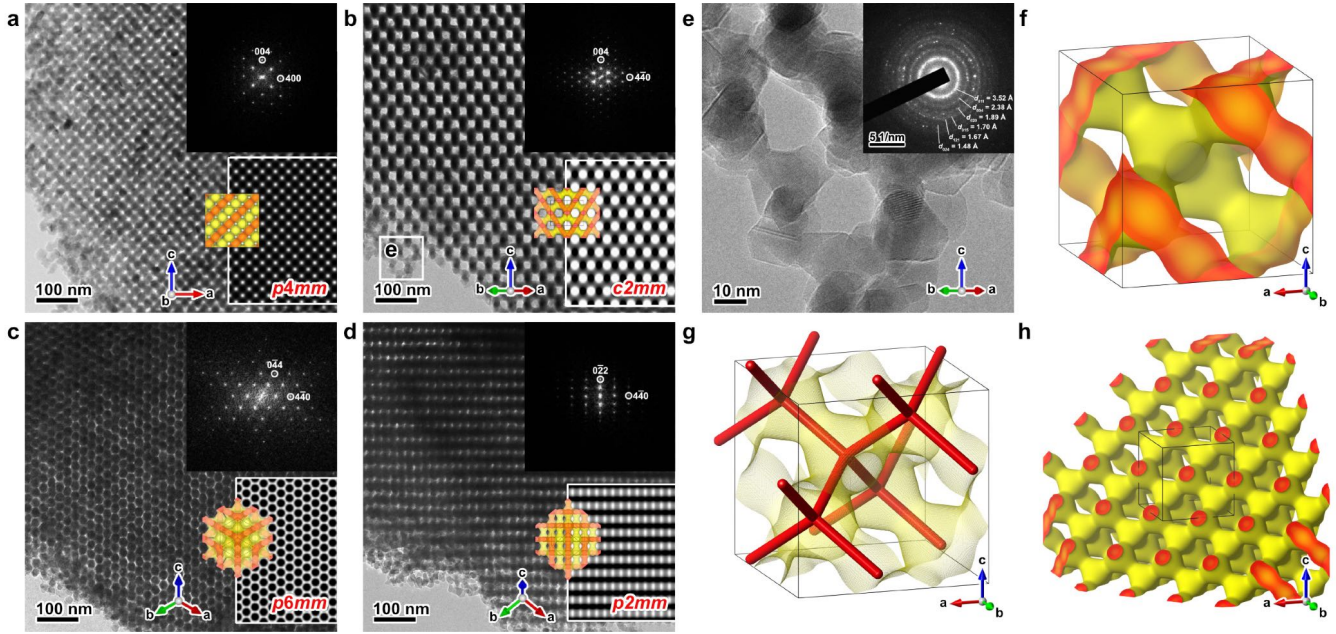

**Fig. 3. Structural characterization of the SD scaffold. a-d**, TEM images and the corresponding FDs of the calcined SD sample taken from the [010], [110], [111] and [112] directions, respectively. The insets show the TEM images simulated with a 3-term nodal equation, and the projections of the reconstructed models were overlaid on both the TEM and simulation. **e**, HRTEM image of the region marked by the white box in **b**, suggesting the polycrystalline nature of anatase TiO$_2$. The inset shows the SAED with the ring pattern. **f**, Reconstructed 3D electrostatic potential map of a unit cell of the SD framework. **g**, The stick model superimposed on the reconstructed volume shows the diamond network with fourfold connectivity. **h**, Cross section of the SD structure from the (111) lattice plane.

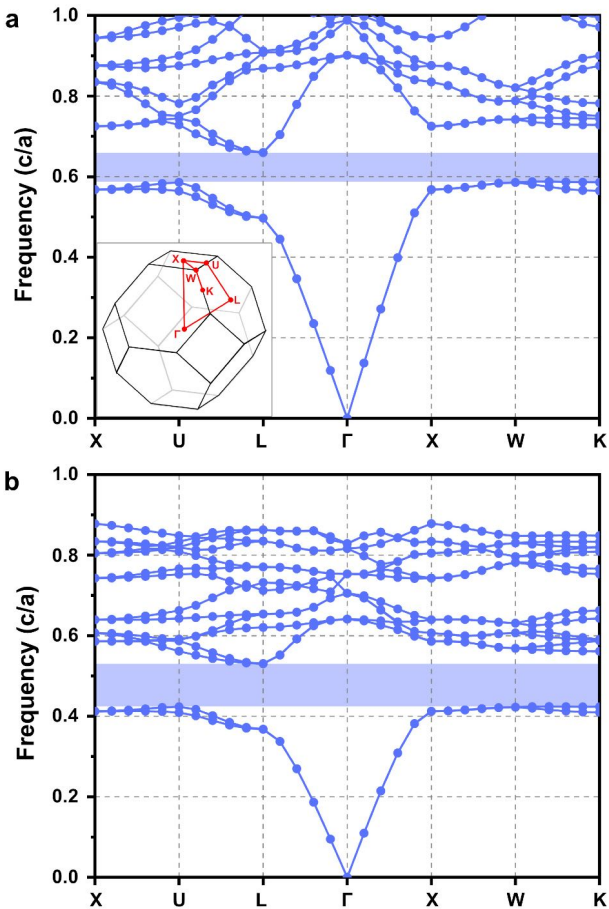

**Fig. 4. Gap map of the SD structure. a**, Gap map of an SD with a dielectric contrast of 6.25:1. **b**, Calculation with a dielectric contrast of 13:1.

The polycrystalline anatase nature of the SD TiO$_2$ scaffold was verified by the selected-area electron diffraction (SAED) pattern and high-resolution TEM (HRTEM) image. The SAED of the whole particle (inset of Fig. 3e) showed a ring pattern, indicating the overall polycrystalline anatase nature. The HRTEM image (Fig. 3e) revealed that the framework comprised stacks of small anatase crystals with sizes of approximately 15-20 nm, which was consistent with the wide-angle XRD analysis.

On the other hand, with an increase in the HCl (3 M)/THF mass ratio from 0.15 to 0.2, accelerated nucleation and condensation of TiO$_2$ led to infiltration of the inorganic precursors into both skeletons of the polymer matrix and resulted in the formation of DD TiO$_2$ frameworks (Supplementary Information Figs. S13 and S14). A comparison of the cross-sectional SEM images (Supplementary Information Fig. S15) clearly indicated that the TiO$_2$ species only occupied one channel in the SD sample but filled both skeletons in the DD structure. With HCl (3 M)/THF contents from 0.25 to 0.375, the polymer matrix was transformed into the cylindrical hexagonal phase with higher interfacial curvature due to the increased volume fraction of the hydrophilic segments caused by H$_2$O. Consequently, TiO$_2$ nanowires originally embedded in the cylindrical pores were obtained after calcination (Supplementary Information Fig. S16).

Propagation of electromagnetic waves in the SD structure was investigated by numerically solving Maxwell's equation with the MIT Photonic-Bands package[36]. The bandstructure was determined by using the reconstructed 3D structural model along the high-symmetry points of the irreducible Brillouin zone with a face-centred cubic lattice. The lower dielectric material was set as air with $\varepsilon_1 = 1$, while the higher dielectric material was assumed to be anatase TiO$_2$ with $\varepsilon_2 = 6.25$. As shown in Fig. 4a, a complete bandgap with a band width of 11.54% was obtained. The gap was increased when the dielectric contrast between the media was enhanced. With a dielectric contrast of $\varepsilon_2 = 13$, the complete bandgap increased to 22.29% (Fig. 4b). Owning to the well-ordered experimental structures, the resulting SD network has promising optical properties and a broad array of potential applications can be envisaged.

Compared to the most appealing gyroid phases, the formation of diamond surfaces is very rare in block copolymeric systems and in other amphiphile self-assembies. It was discovered in a finite area in the phase portrait, which was attributed to the higher free energy than other types of minimal surfaces[37]. Due to the high degree of packing frustration concentrated on the nodal sites of the diamond surface[38], stretching of the molecules had to be relieved or the spaces filled with other substances to stabilize the structure. Methods included the use of copolymer blends with different chain lengths[39], the addition of a homopolymer[40] or solution syntheses[41]. Herein, the diamond minimal surface was formed by $PEO_{45}$-$b$-$PS_{241}$ with a volume fraction of $f_{PEO}$ = 6.9%, providing the proper composition window for the diamond surface[21]. Since we also employed a solution-mediated synthesis, this effectively eliminated packing frustration at the nodal positions. In addition, the inorganic precursors also contributed to the volume fraction of the PEO segment, which was detected at the initial stage of structure formation (Supplementary Information Figs. S17 and S18). These relatively loosely organized hydrophilic oxo oligomers may also have contributed to the interfacial curvature[31]. After solvent evaporation, these oligomers associated with the PEO segments condensed further, leading to the formation of nucleation sites. This nucleation and condensation of the titania species were well controlled by slow evaporation of the solvents and the amount of $H_2O$ and HCl added, which allowed the formation of independent SD domains. Therefore, the above factors caused the packing parameter $p$[21,42] of the assembly system to fall into the phase region favouring formation of the diamond surface structure, and the controlled nucleation and growth of the $TiO_2$ species led to successful formation of the SD configurations.

To the best of our knowledge, this work represents the first report of a bottom-up self-assembled SD structure with a complete photonic bandgap, which has long been sought for photonic structures. This approach solved one of the most challenging problems of self-assembly with a very straightforward and universal approach, one that can also be extended to create other unbalanced biological structures that are difficult to achieve through conventional template syntheses. Owing to the highly flexible and versatile synthetic behaviour of block copolymer self-assembly, various novel photonic materials with controllable structures and adjustable lattice parameters can be expected[18,28]. This work provides significant opportunities in various research areas and opens new horizons for studies of self-assembly, the development of new biologically relevant materials and next-generation optical devices, information processing, and energy technologies.

## Acknowledgements

The authors thank Prof. Hexiang Deng and Mr. Gaoli Hu, Wuhan University, for the support for SAXS measurements. This work was financially supported by the National Natural Science Foundation of China (Grant Nos. 21922304, L.H., 22225501, Y.M., 22005096, Y.C.) and Fundamental Research Funds for the Central Universities (L.H.).

## Author Contributions

L.H. conceived the idea and led the project. C.W. synthesized the materials and performed the GPC, NMR, SAXS, XRD and SEM measurements of the samples. L.H. performed the TEM observation and 3D reconstruction. Q.D. contributed to the structural analysis. C.C. calculated the photonic bandgap. S.A. performed the cross-sectional SEM observation and EDS analysis. S.C., C.Z. Y.C. and Y.M. participated in discussions. All authors participated in the preparation of the manuscript.

## Competing interests

The authors declare no conflict of interest.

# Supporting Information

# Single Diamond Structured Titania Scaffold

Chao Wang[1], Congcong Cui[1], Quanzheng Deng[1], Chong Zhang[1], Shunsuke Asahina[2], Yuanyuan Cao[3], Yiyong Mai[4], Shunai Che[1,4], Lu Han[1]*

[1]School of Chemical Science and Engineering, Tongji University, 1239 Siping Road, Shanghai, 200092, China.
[2]SMBU, JEOL, Akishima, Tokyo 196-8558, Japan.
[3]School of Materials Science and Engineering, East China University of Science and Technology; Shanghai, 200237, China.
[4] School of Chemistry and Chemical Engineering, State Key Laboratory of Composite Materials, Shanghai Key Laboratory for Molecular Engineering of Chiral Drugs, Shanghai Jiao Tong University, 800 Dongchuan Road, Shanghai, 200240, China.
*e-mail: luhan@tongji.edu.cn

**Experimental Section**

**Chemicals**

Titanium diisopropoxide bis(acetylacetonate) (TIA, 75% in isopropyl alcohol, Adamas), 2-bromoisobutyryl bromide (98%, J&K Scientific), 1,1,4,7,7-pentamethyldiethylenetriamine (PMDETA, 99%, TCI), triethylamine (99%, J&K Scientific), 4-dimethylaminopyridine (99%, J&K Scientific), copper bromide (CuBr, 98.5%, Sinopharm Chemical Reagent Co., Ltd), anhydrous diethyl ether (99%, Sinopharm Chemical Reagent Co. Ltd), hydrochloric acid (HCl, 36%~38%, Sinopharm Chemical Reagent Co., Ltd), methanol (99.7%, Sinopharm Chemical Reagent Co., Ltd), tetrahydrofuran (THF, 99.8%, Sinopharm Chemical Reagent Co., Ltd), dichloromethane ($CH_2Cl_2$, 99.5%, Sinopharm Chemical Reagent Co., Ltd), *N,N*-dimethylformamide (DMF, 99.9%, Alfa Aesar) and deionized water (Milli-Q, 18.2 MΩ·cm) were used without further purification. Styrene (99.5%, Sinopharm Chemical Reagent Co., Ltd), contained a polymerization inhibitor that was removed with a 20% sodium hydroxide solution; the styrene was dried with anhydrous magnesium sulfate and then distilled under reduced pressure before use. Methyl poly(ethylene oxide) with a hydroxyl terminal group (PEO-OH 2000) was purchased from Aldrich and labelled $PEO_{45}$.

**Synthesis of the block copolymer PEO$_{45}$-*b*-PS$_{241}$**

PEO$_{45}$-Br was prepared through the esterification of PEO$_{45}$-OH with 2-bromoisobutyryl bromide in CH$_2$Cl$_2$. PEO$_{45}$-OH (40.0 g, 20.0 mmol) was dissolved in dried CH$_2$Cl$_2$ (200 mL). To the solution were added 2.5 mL of triethylamine and 3.0 g of 4-dimethylaminopyridine. After the solution was stirred for 30 min while cooled in an ice water bath, 2-bromoisobutyryl bromide (23.0 g, 100.0 mmol) was added to the solution over a period of 1 h under N$_2$ protection. Finally, the solution was stirred at room temperature for 12 h and then filtered to obtain a homogeneous solution. Subsequently, 500 mL of cold ether was added to the solution to obtain a white precipitate. The precipitate was filtered and washed several times with cold ether. After the precipitate was dried under vacuum, the macroinitiator PEO$_{45}$-Br was obtained. $^1$H NMR (CDCl$_3$), δ(ppm): 1.90 (s, 6H, -C(CH$_3$)$_2$Br), 3.50-3.70 (m, 4H, -CH$_2$CH$_2$-O- of the PEO chain).

PEO$_{45}$-*b*-PS$_{241}$ was prepared via atom-transfer radical polymerization (ATRP) of styrene using PEO$_{45}$-Br as the initiator and the CuBr/ PMDETA catalyst system. A 250 mL Schlenk flask containing 1.0 g of PEO$_{45}$-Br (0.4 mmol), 180 μL of PMDETA (0.9 mmol), and 10.0 g of styrene (96.2 mmol) was thoroughly purged and then sealed with a rubber stopper. After the solution became clear while stirring, 0.045 g of CuBr (0.3 mmol) was added to the solution. The flask containing the reactants was fully degassed with a minimum of four freeze-pump-thaw cycles and sealed under vacuum. The flask was subsequently placed in a 110 °C oil bath to allow the polymerization to proceed. Several hours later, the reaction mixture was cooled to room temperature. The polymerization was terminated by exposing the reaction mixture to air. The catalyst was removed by filtration through neutral alumina using CH$_2$Cl$_2$ as the eluent. The polymer was precipitated with cold methanol (500 mL) and was subsequently dried under vacuum. The $M_{n,GPC}$ of PEO$_{45}$-*b*-PS was approximately 2.71 × 10$^4$ g/mol, with a Đ of 1.12. The degree of polymerization for the polystyrene segment was determined to be 241. $^1$H NMR (CDCl$_3$), δ(ppm): 3.50–3.70 (m, 4H, -CH$_2$CH$_2$-O- of the PEO chain), 6.30–7.20 (m, 5H, -C$_6$H$_5$ of the PS chain).

**Synthesis of the SD TiO$_2$ scaffold**

The TiO$_2$ scaffold was prepared via solvent evaporation and the use of the diblock copolymer PEO$_{45}$-*b*-PS$_{241}$ as the template. A total of 0.05 g of the template was dissolved in 8.0 g of THF. After stirring for approximately 2 h, 0.8 g of HCl (3 M) was added to the above solution. After stirring for 2 h, 0.5 g of TIA was added to the mixture. The solution was maintained under stirring for an additional 2 h. The solution was then evaporated at ambient temperature to remove the volatile components. The as-synthesized samples were calcined under air at 500 °C for 6 h.

**Characterization**

The molecular weights and polydispersity index (PDI) were determined with an HLC-8420GPC (TOSOH Corp.) gel permeation chromatography (GPC) apparatus. The measurements were conducted at a flow rate of 500 μL/min using DMF as the eluent. Nuclear magnetic resonance (NMR) spectra were measured on a Varian Mercury Plus 600 MHz NMR spectrometer using tetramethylsilane (TMS) as the internal reference, and the polymers were dissolved in deuterated chloroform.

The SAXS patterns were collected on a Rigaku NANOPIX system. The X-ray generator was Fr-x (rotation anode X-ray generator) with a power of 2.97 kw (45 kV, 66 mA) and a target/wavelength of Cu/1.5418 Å. The camera length was 1290 mm. The incident beam was in New CMF optics, and the detector was a HyPix-54000. All 1D profiles were collected with two pinholes and high-resolution collimation with a beam stop size of 2.0 mmφ. Before the measurements, the camera length was calibrated with silver behenate.

The XRD patterns were recorded with a Rigaku MiniFlex 600 powder diffractometer equipped with Cu Kα radiation (40 kV, 15 mA) at a scan rate of 0.2° min$^{-1}$ over the range 10-90°.

Cross-sectioning was performed by Argon ion beam polishing using a JEOL IB-19520 cross-section polisher with an acceleration voltage of 4 kV. The sample was cooled by liquid nitrogen to -50°C during the 5 h polishing process.

SEM images were obtained with a JEOL JSM-7900F system operated at a low accelerating voltage for the landing energy of 1 keV with a specimen bias of -5 kV (point resolution of 0.7 nm). For SEM observations, the bulk samples were lightly crushed into pieces in an agate mortar and dispersed onto adhesive conductive glue fixed on a Cu sample holder without any metal coating.

The observations of the cross sections were carried out using a JEOL JSM-IT800<i> with a semi-in-lens-type objective lens. The EDS measurements were performed with an Oxford Extreme 100 mm$^2$ windowless detector with an accelerating voltage of 8 kV, a probe current of 100 pA and a mapping time of 10 min.

The TEM experiments were performed with a JEOL JEM-F200 microscope equipped with a Schottky gun operated at 200 kV (Cs 1.0 mm, Cc 1.1 mm, point resolution of 1.9 Å for TEM mode). The images were recorded with a GATAN OneView IS camera (4096 × 4096 pixels) under low-dose conditions. For the TEM observations, the samples were crushed, dispersed in ethanol and dripped onto a carbon thin film on a Cu grid. The software MesoPoreImage was used for TEM image simulation[35], which calculated the projected potential and the TEM images from a 3D continuum model employing a mathematical approach of the single diamond surface structure using nodal approximation[36]. The threshold value for the simulation was -3.8 < $t$ < 3.8 for the diamond surface and $t$ > 2.1 for the SD network. The simulation conditions were an accelerating voltage of 200 kV, Cs of 1.0 mm, Cc of 1.1 mm, sample thickness of 200 nm and defocus value of 300 nm.

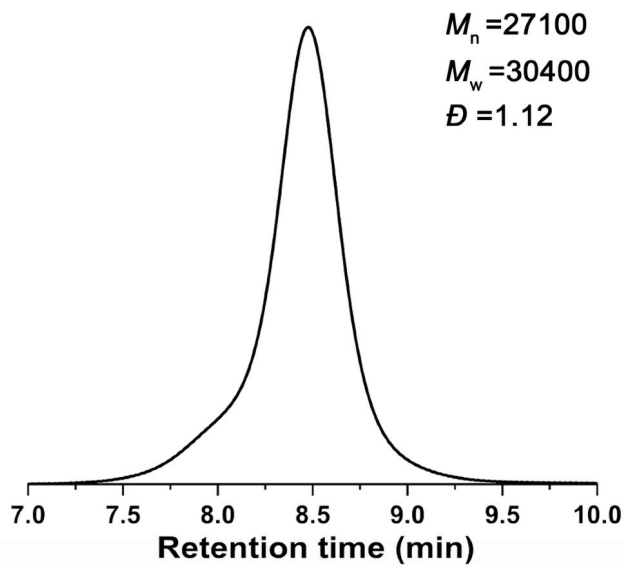

**Fig. S1. Gel permeation chromatogram of the PEO$_{45}$-*b*-PS$_{241}$ block copolymer.**

Fig. S1 shows the GPC curve for the PEO$_{45}$-*b*-PS$_{241}$ block copolymer. The degree of polymerization (DP) for the PS block was calculated to be 241. The molecular weight dispersity index (*Ð*) of 1.12 suggested a narrow molecular weight distribution.

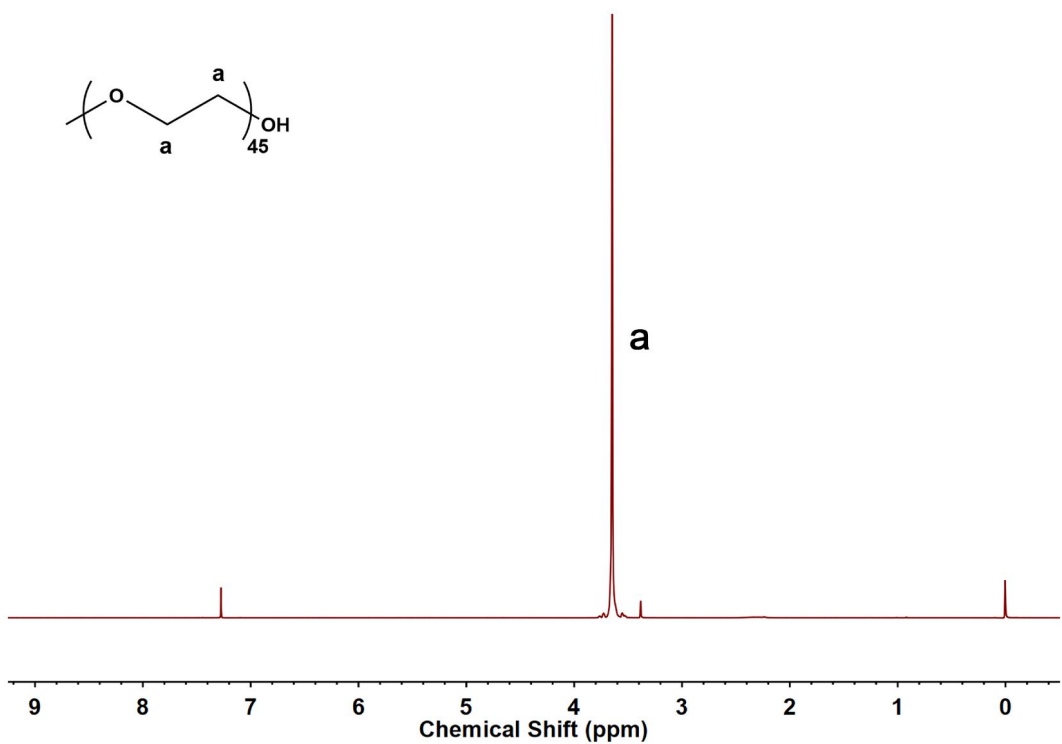

**Fig. S2. ¹H NMR spectrum of PEO$_{45}$-OH.** The polymer was dissolved in deuterated chloroform.

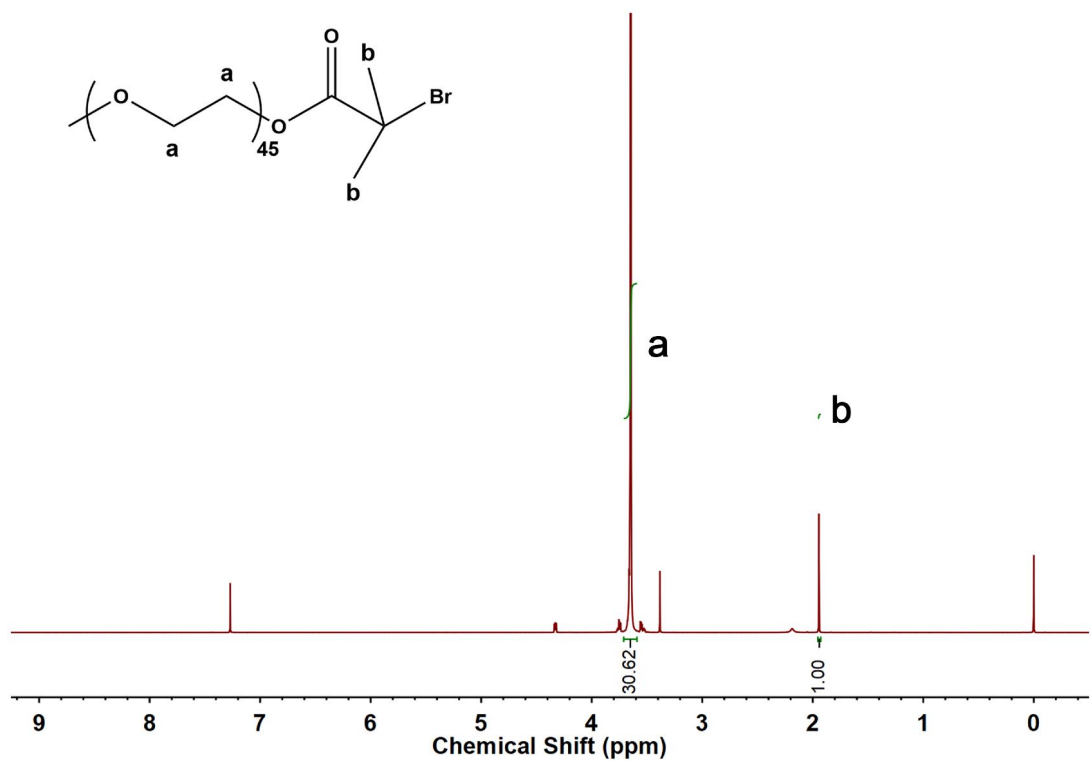

Fig. S3. $^1$H NMR spectrum of PEO$_{45}$-Br. The polymer was dissolved in deuterated chloroform.

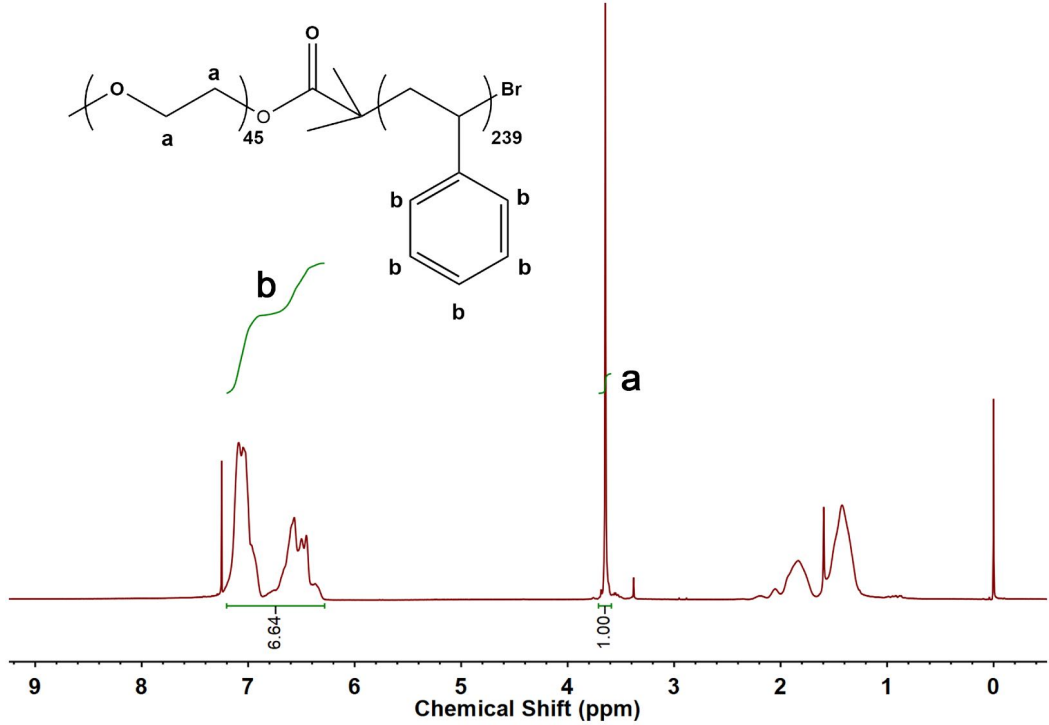

**Fig. S4. $^1$H NMR spectrum of PEO$_{45}$-b-PS$_{241}$.** The polymer was dissolved in deuterated chloroform.

Fig. S4 shows the $^1$H NMR spectrum of the PEO$_{45}$-b-PS$_{241}$ copolymer (241 is the DP of the PS block calculated from the GPC result). From the $^1$H NMR spectrum, the DP of PS was calculated as follows:

$$DP_{PS} = \frac{I_b/5}{I_a/4} \times 45 = \sim 239$$

where $I_a$ is the integrated value of the proton peaks attributed to the PEO block (signal a) and $I_b$ represents the integrated value of the proton peaks attributed to the PS block (signal b). 45 was the DP of the PEO block. The DP value calculated from the NMR spectra was in good agreement with the DP value calculated from the GPC data.

The volume fraction ($f_{PS}$) of the PS block was calculated as follows:

$$f_{PS} = \frac{V_{PS}}{V_{PS}+V_{PEO}} = \frac{M_{PS}/\rho_{PS}}{M_{PS}/\rho_{PS}+M_{PEO}/\rho_{PEO}} = \frac{25100/1.05}{25100/1.05+2000/1.13} = \sim 93.1\%$$

where $V_{PEO}$ and $V_{PS}$ are the volumes of PEO and the PS coils, respectively; $M_{PEO}$ and $M_{PS}$ are the molecular weights of the PEO and PS blocks, respectively. $\rho_{PEO}$ is the density of PEO, which is approximately 1.13 g/cm$^3$, and $\rho_{PS}$ is the density of PS (1.05 g/cm$^3$)[1].

**Table S1.** Number-averaged molecular weights, densities and volume fractions of different segments.

| Component | $M_n$ (kg/mol) | $\rho$ (g/cm³) | Vol% | Đ |
|---|---|---|---|---|
| PEO | 2.0 | 1.13 | 6.9 | |
| PS | 25.1 | 1.05 | 93.1 | |
| PEO-*b*-PS | 27.1 | | | 1.12 |

**Table S2.** Solubility parameters of different blocks.

| Blocks | $\delta_d$ (J/cm³)$^{1/2}$ | $\delta_d$ (J/cm³)$^{1/2}$ | $\delta_h$ (J/cm³)$^{1/2}$ | $\delta$ (J/cm³)$^{1/2}$ |
|---|---|---|---|---|
| PEO | 16.7 | 10.1 | 8.1 | 21.2 |
| PS | 16.7 | 8.3 | 5.1 | 19.3 |

The Flory–Huggins interaction parameters for the copolymer were calculated as follows:

$$\delta^2 = \delta_d^2 + \delta_p^2 + \delta_h^2$$

$$\chi N = \sum N \times V_{ref}(\delta_1 - \delta_2)^2 / RT$$

where $V_{ref}$ is the segment reference volume (100 cm³/mol) and $\delta_i$ is the Hildebrand solubility parameter for polymer i (Table S2).

**Table S3.** Flory–Huggins interaction parameter between blocks.

| Blocks | PEO-PS |
|---|---|
| $\chi N$ | 43.1 |

The Flory-Huggins interaction parameter of $\chi N \approx 43.1$ suggested strong microphase separation.

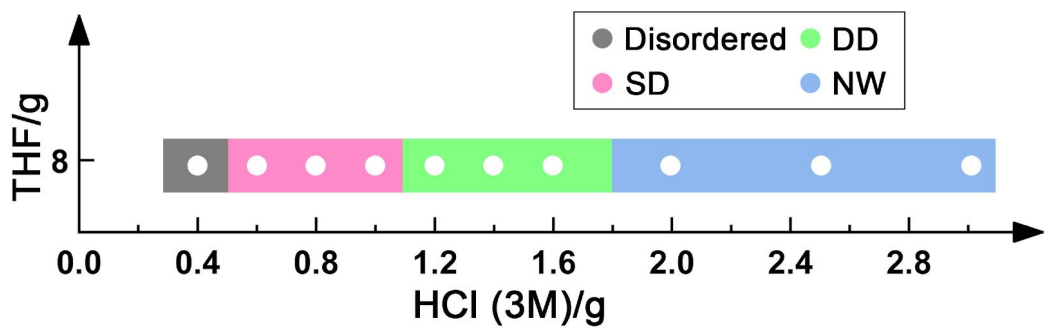

**Fig. S5. Synthesis-field diagram.** The synthetic mixture included 0.05 g of PEO$_{45}$-*b*-PS$_{241}$, 8.0 g of THF, *X* g of HCl (3 M) and 0.5 g of TIA. Various structures, including disordered, single diamond (SD), double diamond (DD) and nanowire (NM), were synthesized from the PEO$_{45}$-*b*-PS$_{241}$ template by varying the amount of HCl (3 M).

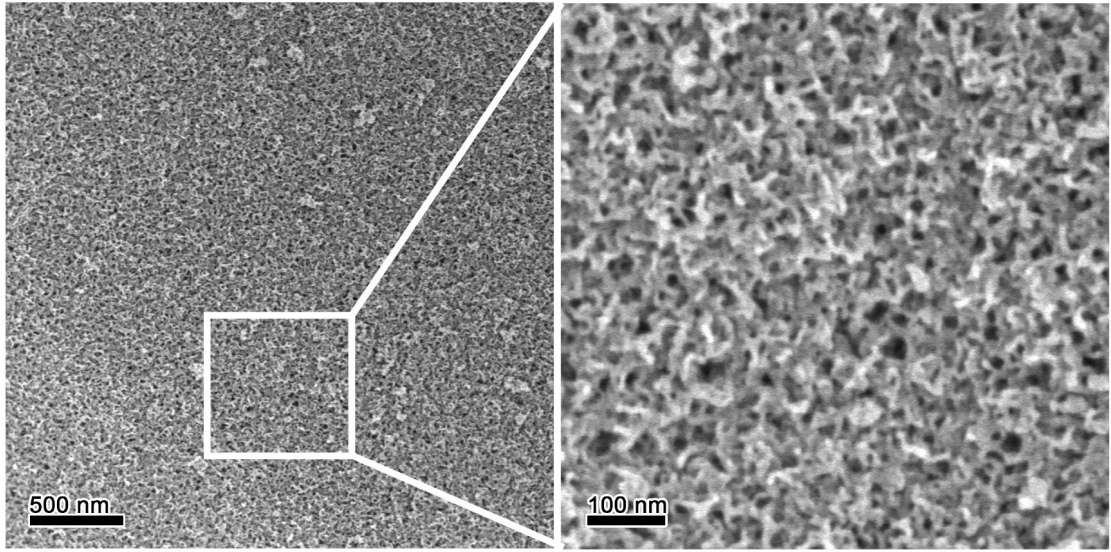

**Fig. S6. SEM images of the disordered structure.** The synthetic mixture included 0.05 g of PEO$_{45}$-*b*-PS$_{241}$, 8.0 g of THF, 0.4 g of HCl (3 M) and 0.5 g of TIA.

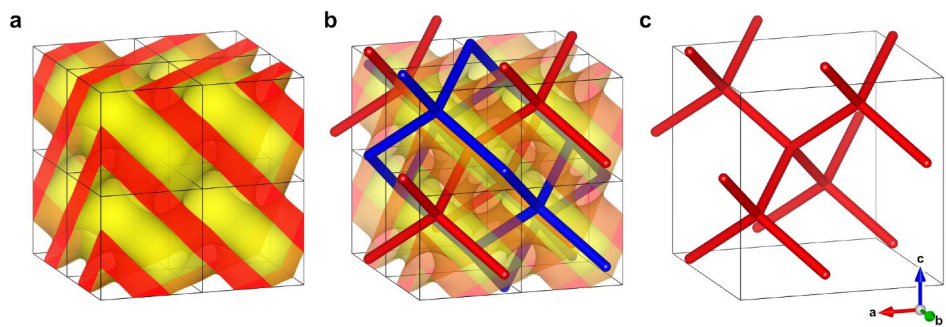

**Fig. S7. Geometric relationship between the diamond minimal surface and the single diamond network. a**, Schematic drawing of the diamond minimal surface with the space group $Pn\bar{3}m$ (2 × 2 × 2 unit cells). **b**, The stick model indicates interpenetrated double diamond networks separated by the diamond minimal surface. **c**, The single diamond network with space group $Fd\bar{3}m$. The unit cell parameter is twice that of the diamond minimal surface.

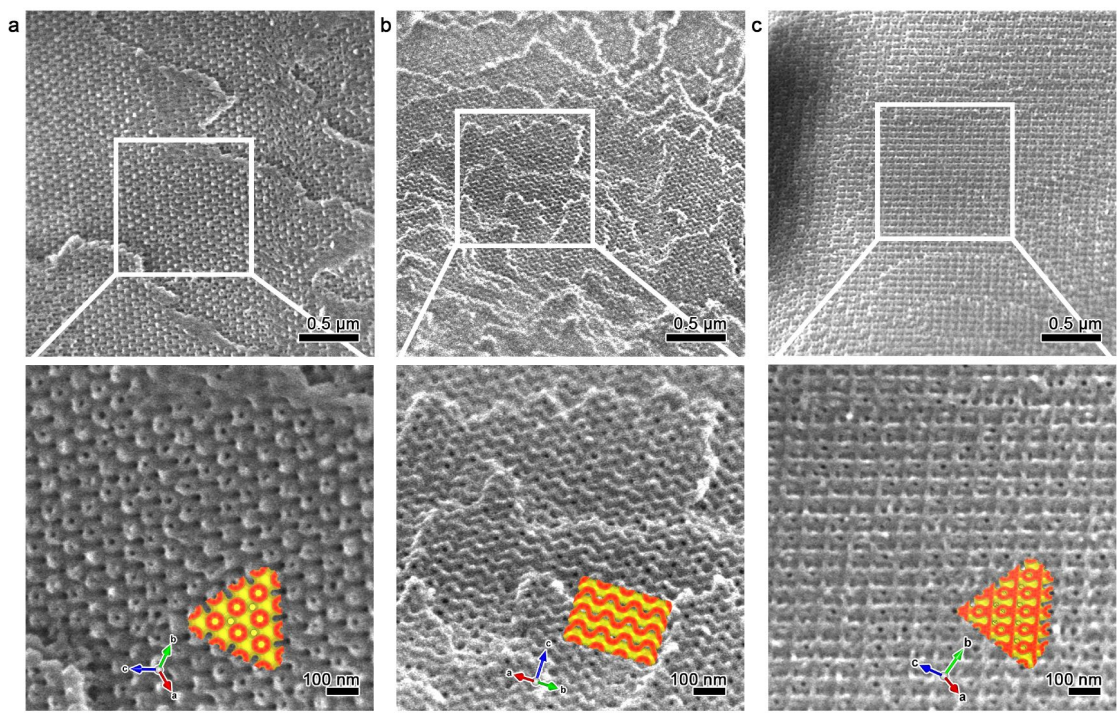

**Fig. S8. SEM images of the as-synthesized SD sample. a-c**, SEM images of the as-synthesized SD sample taken from the {111}, {110} and {112} lattice planes, respectively. The synthetic mixture included 0.05 g of PEO$_{45}$-*b*-PS$_{241}$, 8.0 g of THF, 0.8 g of HCl (3 M) and 0.5 g of TIA.

Fig. S9 reveals the TEM images and the corresponding Fourier diffractograms (FDs) of the as-synthesized sample taken from the thin edge (polymer matrix part) along the [110], [111], [112] and [010] directions. The extinction conditions from the FDs can be summarized as {0$kl$: $k + l$ = 2n; 00$l$: $l$ = 2n; $hkl$: none and $hhl$: none}, suggesting two possible space groups, $Pn\bar{3}$ (No. 201) and $Pn\bar{3}m$ (No. 224). The symmetry-averaged TEM images were generated by the crystallographic image processing software CRISP[2], in which the two-dimensional (2D) plane group symmetries of *p2mm*, *p6mm*, *p2mm* and *p4mm* were justified. The space group $Pn\bar{3}m$ was uniquely chosen due to the plane group symmetry from the [111] axis. The unit cell parameter $a \approx$ 45 nm calculated from the FDs was consistent with the SAXS analysis.

Electron crystallography three-dimensional (3D) reconstruction was applied to reveal the configurations of the samples. The structure factor amplitudes and phases were extracted from the FDs taken from the corresponding TEM images. All the structure factors corresponding to each 2D projection were then adjusted to the common crystal origin. The structure factors were combined into one dataset upon normalization by the common reflection intensities (Table S4). Then, the 3D electrostatic potential distributions $\varphi(x,y,z)$ were calculated by inverse Fourier transform employing the software VESTA[3] after the correction for the contrast transfer function using a Wiener filter. The threshold value for the equi-electrostatic potential surface was determined from the TEM images. In the reconstructed structure, the typical diamond minimal surface is readily recognized.

TEM images were simulated with the software MesoPoreImage[4] to verify the structural solution. The simulations from different zone axes are shown as insets in the TEM images, and they were highly consistent with the observed TEM contrast.

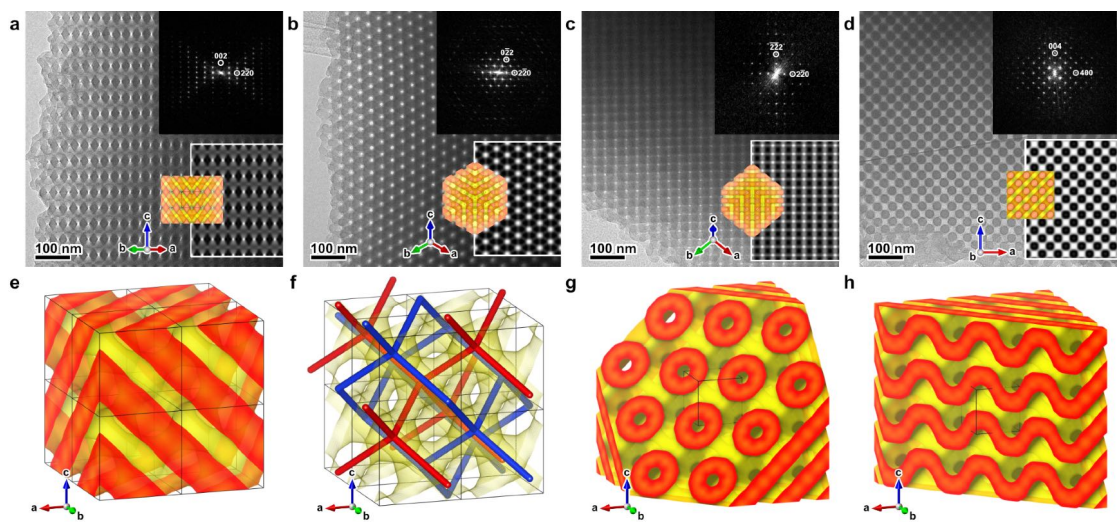

**Fig. S9. Structural characterizations of the polymer matrix of the as-synthesized sample. a-d**, TEM images and the corresponding FDs of the as-synthesized sample taken from the thin edge of the polymer template. The images were taken from the [110], [111], [112] and [100] directions. The insets show TEM images simulated with a 3-term nodal equation, and the projections of the reconstructed models are overlaid on both the TEM and simulation. **e**, Reconstructed 3D electrostatic potential map of 2 × 2 × 2 unit cells. **f**, The stick model shows the double skeleton separated by the diamond minimal surface polymer matrix. **g**, **h**, Cross sections of the reconstructed structure from the (111) and (110) lattice planes, respectively. The synthetic mixture included 0.05 g of $PEO_{45}$-*b*-$PS_{241}$, 8.0 g of THF, 0.8 g of HCl (3 M) and 0.5 g of TIA.

**Table S4. Crystal structure factors (amplitudes and phases) extracted from the TEM images of the polymer matrix.**

| h | k | l | Amplitude | Phase | h | k | l | Amplitude | Phase |
|---|---|---|-----------|-------|---|---|---|-----------|-------|
| 1 | 1 | 0 | 10000 | 180 | 7 | 3 | 1 | 9 | 0 |
| 1 | 1 | 1 | 4639 | 0 | 6 | 5 | 1 | 9 | 180 |
| 2 | 0 | 0 | 1114 | 180 | 7 | 3 | 2 | 8 | 180 |
| 2 | 1 | 1 | 1227 | 0 | 5 | 5 | 4 | 18 | 180 |
| 2 | 2 | 0 | 1404 | 0 | 7 | 4 | 1 | 13 | 0 |
| 2 | 2 | 1 | 1268 | 180 | 6 | 4 | 4 | 11 | 0 |
| 3 | 1 | 0 | 464 | 180 | 8 | 2 | 0 | 7 | 180 |
| 3 | 1 | 1 | 333 | 0 | 6 | 6 | 0 | 10 | 180 |
| 2 | 2 | 2 | 769 | 0 | 8 | 2 | 2 | 6 | 0 |
| 3 | 2 | 1 | 260 | 0 | 6 | 6 | 1 | 9 | 180 |
| 4 | 0 | 0 | 199 | 180 | 7 | 4 | 3 | 15 | 0 |
| 3 | 2 | 2 | 54 | 180 | 7 | 5 | 0 | 5 | 180 |
| 3 | 3 | 0 | 220 | 180 | 5 | 5 | 5 | 21 | 0 |
| 4 | 1 | 1 | 135 | 0 | 7 | 5 | 1 | 16 | 180 |
| 3 | 3 | 1 | 263 | 0 | 6 | 6 | 2 | 8 | 0 |
| 4 | 2 | 0 | 237 | 0 | 6 | 5 | 4 | 17 | 0 |
| 4 | 2 | 1 | 241 | 180 | 8 | 3 | 2 | 12 | 0 |
| 3 | 3 | 2 | 256 | 180 | 7 | 5 | 2 | 11 | 0 |
| 4 | 2 | 2 | 48 | 0 | 8 | 4 | 0 | 5 | 180 |
| 4 | 3 | 1 | 30 | 0 | 6 | 6 | 3 | 9 | 180 |
| 5 | 1 | 0 | 33 | 0 | 7 | 4 | 4 | 9 | 180 |
| 3 | 3 | 3 | 234 | 0 | 7 | 5 | 3 | 6 | 0 |
| 5 | 1 | 1 | 30 | 0 | 8 | 4 | 2 | 4 | 180 |
| 4 | 3 | 2 | 150 | 0 | 6 | 5 | 5 | 11 | 180 |
| 5 | 2 | 1 | 40 | 180 | 7 | 6 | 1 | 6 | 0 |
| 4 | 4 | 0 | 39 | 0 | 6 | 6 | 4 | 8 | 0 |
| 4 | 4 | 1 | 70 | 180 | 9 | 3 | 0 | 4 | 180 |
| 4 | 3 | 3 | 60 | 180 | 8 | 5 | 2 | 3 | 0 |
| 5 | 3 | 0 | 53 | 180 | 8 | 4 | 4 | 5 | 180 |
| 5 | 3 | 1 | 64 | 0 | 6 | 6 | 5 | 8 | 180 |
| 4 | 4 | 2 | 66 | 0 | 7 | 7 | 0 | 5 | 0 |
| 6 | 0 | 0 | 39 | 180 | 8 | 5 | 3 | 5 | 180 |
| 5 | 3 | 2 | 21 | 180 | 9 | 4 | 1 | 4 | 0 |
| 6 | 1 | 1 | 34 | 0 | 7 | 5 | 5 | 5 | 0 |
| 6 | 2 | 0 | 41 | 0 | 7 | 7 | 0 | 6 | 0 |
| 4 | 4 | 3 | 58 | 180 | 9 | 3 | 3 | 6 | 180 |
| 5 | 3 | 3 | 15 | 0 | 8 | 6 | 1 | 9 | 0 |
| 6 | 2 | 2 | 9 | 180 | 8 | 6 | 2 | 5 | 180 |
| 4 | 4 | 4 | 76 | 0 | 10 | 2 | 0 | 4 | 180 |
| 5 | 4 | 3 | 52 | 0 | 7 | 7 | 3 | 3 | 0 |
| 5 | 5 | 0 | 16 | 180 | 6 | 6 | 6 | 5 | 0 |
| 7 | 1 | 0 | 19 | 0 | 7 | 6 | 5 | 5 | 0 |
| 5 | 5 | 1 | 23 | 0 | 9 | 5 | 2 | 4 | 0 |
| 6 | 4 | 0 | 8 | 0 | 7 | 7 | 4 | 4 | 180 |
| 6 | 4 | 1 | 12 | 180 | 8 | 6 | 4 | 3 | 0 |
| 6 | 3 | 3 | 13 | 180 | 7 | 6 | 6 | 5 | 180 |
| 5 | 5 | 2 | 25 | 180 | 9 | 6 | 3 | 3 | 180 |
| 6 | 4 | 2 | 15 | 0 | 7 | 7 | 5 | 3 | 0 |
| 5 | 4 | 4 | 28 | 180 | 8 | 8 | 0 | 3 | 180 |
| 7 | 3 | 0 | 18 | 180 | 9 | 7 | 1 | 3 | 180 |
| 5 | 5 | 3 | 21 | 0 | 7 | 7 | 6 | 3 | 180 |

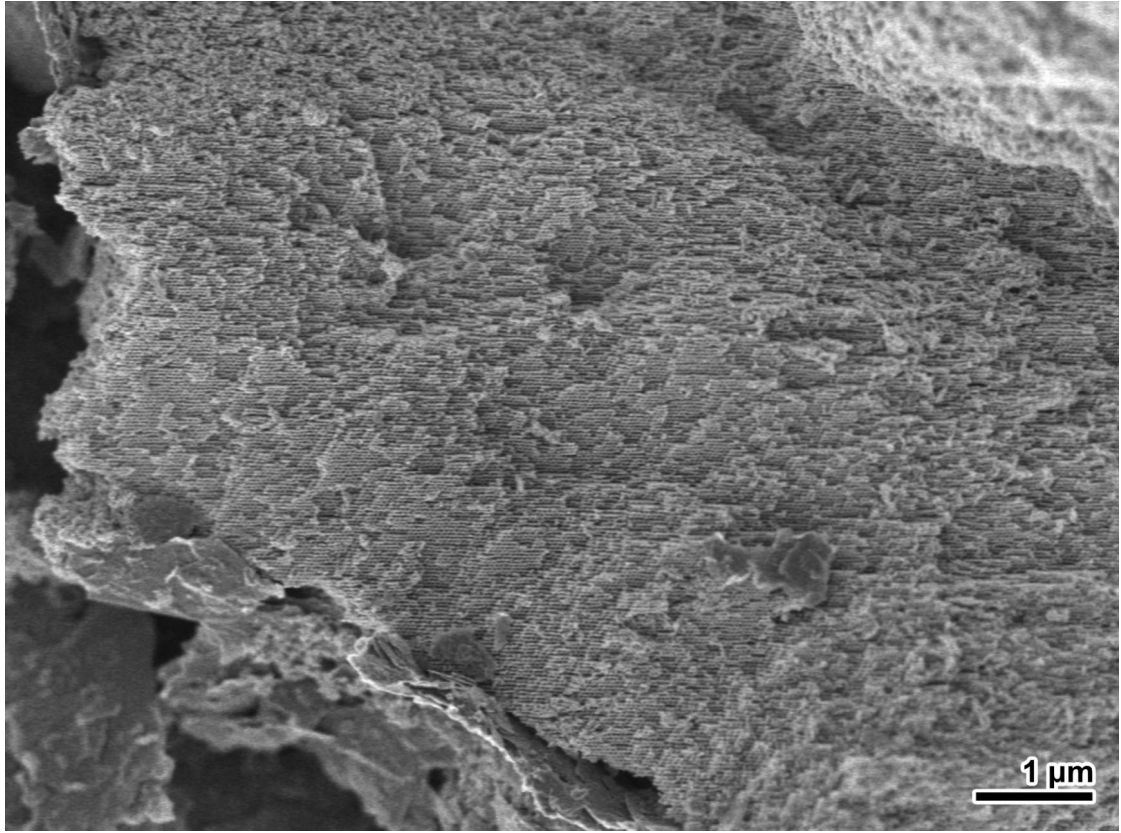

**Fig. S10. Low magnification SEM image of the calcined SD sample.** The single crystal SD domain reached more than 10 μm in size. The synthetic mixture included 0.05 g of PEO$_{45}$-*b*-PS$_{241}$, 8.0 g of THF, 0.8 g of HCl (3 M) and 0.5 g of TIA.

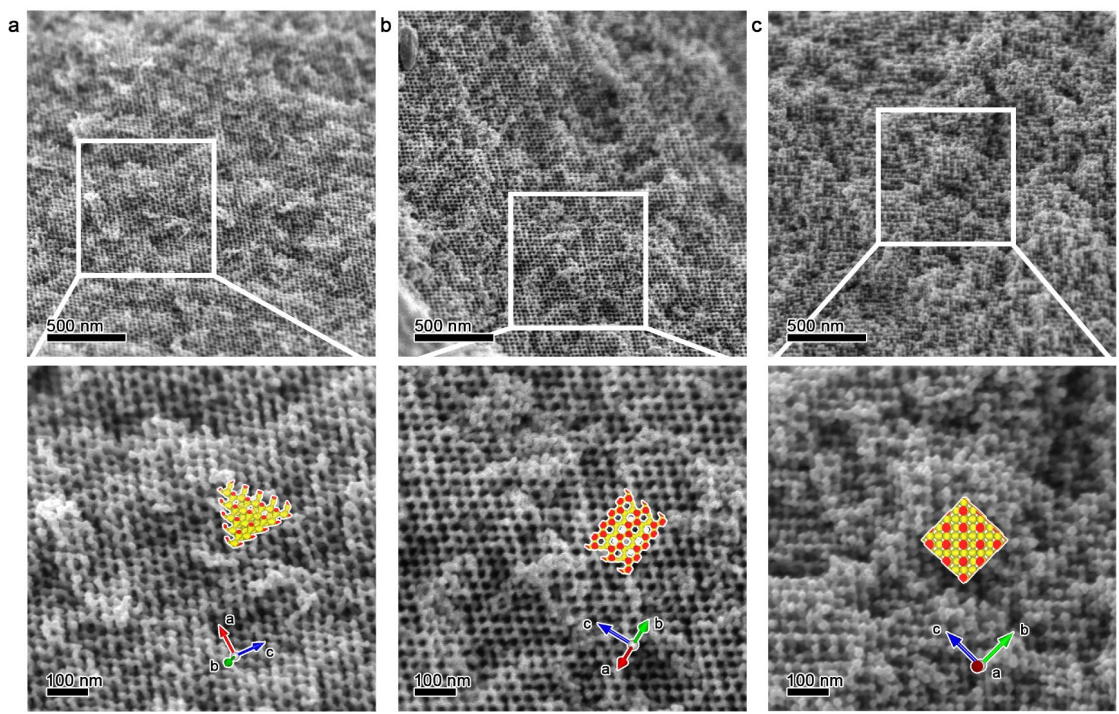

**Fig. S11. Low magnification SEM images of the calcined SD TiO₂ scaffold. a-c**, SEM images of the calcined product taken near the [183], [110] and [001] directions, respectively. The synthetic mixture included 0.05 g of PEO$_{45}$-*b*-PS$_{241}$, 8.0 g of THF, 0.8 g of HCl (3 M) and 0.5 g of TIA.

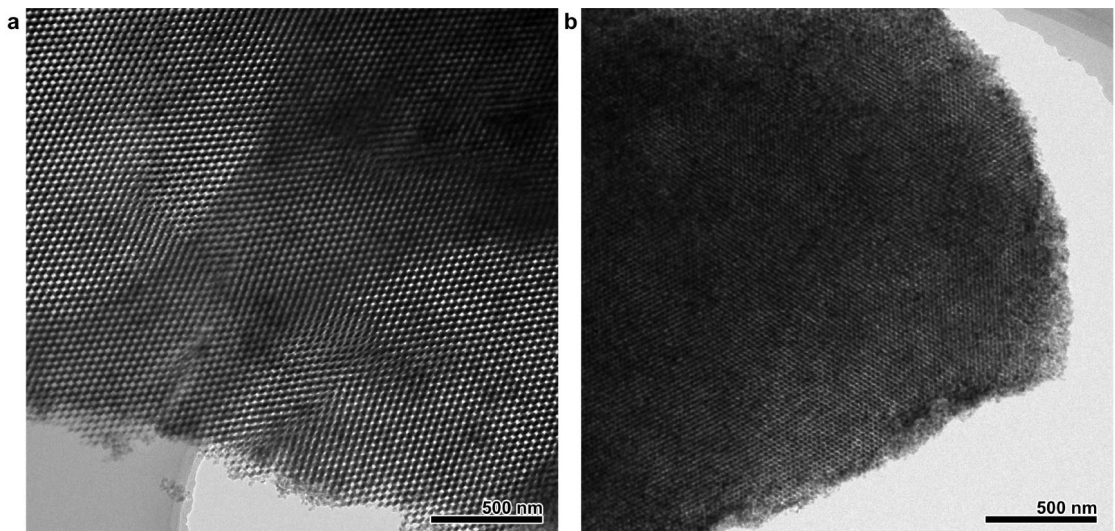

**Fig. S12. Low magnification TEM image of the calcined SD sample. a**, [110] axis. **b**, [111] axis. The synthetic mixture included 0.05 g of PEO$_{45}$-*b*-PS$_{241}$, 8.0 g of THF, 0.8 g of HCl (3 M) and 0.5 g of TIA.

**Table S5.** Crystal structure factors (amplitudes and phases) extracted from the TEM images of the calcined SD sample.

| h | k | l | Amplitude | Phase | h | k | l | Amplitude | Phase |
|---|---|---|---|---|---|---|---|---|---|
| 1 | 1 | 1 | 10000 | 180 | 0 | 0 | 8 | 37 | 180 |
| 2 | 2 | 0 | 1732 | 0 | 0 | 2 | 8 | 33 | 0 |
| 3 | 1 | 1 | 391 | 0 | 2 | 5 | 7 | 28 | 168 |
| 2 | 2 | 2 | 369 | 0 | 0 | 6 | 6 | 28 | 0 |
| 4 | 0 | 0 | 442 | 180 | 3 | 4 | 7 | 26 | 116 |
| 3 | 3 | 1 | 442 | 180 | 2 | 2 | 8 | 26 | 0 |
| 4 | 2 | 2 | 263 | 0 | 2 | 4 | 4 | 49 | 90 |
| 3 | 3 | 3 | 42 | 0 | 2 | 2 | 6 | 34 | 270 |
| 5 | 1 | 1 | 147 | 0 | 2 | 5 | 5 | 27 | 180 |
| 4 | 4 | 0 | 106 | 0 | 0 | 7 | 7 | 14 | 90 |
| 5 | 3 | 1 | 32 | 0 | 0 | 5 | 5 | 28 | 270 |
| 4 | 4 | 2 | 55 | 0 | 1 | 1 | 8 | 19 | 90 |
| 6 | 2 | 0 | 53 | 180 | 1 | 6 | 7 | 14 | 224 |
| 6 | 2 | 2 | 21 | 180 | 0 | 1 | 7 | 24 | 90 |
| 4 | 4 | 4 | 13 | 0 | 1 | 5 | 6 | 19 | 271 |
| 5 | 5 | 1 | 48 | 180 | 6 | 6 | 6 | 10 | 270 |
| 7 | 1 | 1 | 16 | 0 | 0 | 5 | 9 | 10 | 270 |
| 6 | 4 | 2 | 72 | 0 | 1 | 7 | 8 | 8 | 324 |
| 5 | 5 | 3 | 15 | 0 | 0 | 2 | 10 | 9 | 180 |
| 8 | 0 | 0 | 26 | 0 | 0 | 3 | 9 | 10 | 90 |
| 6 | 6 | 0 | 15 | 0 | 1 | 1 | 10 | 9 | 180 |
| 8 | 2 | 2 | 10 | 180 | 0 | 8 | 8 | 7 | 0 |
| 7 | 5 | 1 | 5 | 0 | 0 | 6 | 8 | 8 | 180 |
| 6 | 6 | 2 | 11 | 0 | 4 | 5 | 9 | 7 | 109 |
| 8 | 4 | 0 | 4 | 180 | 2 | 6 | 6 | 10 | 90 |

Fig. S13 shows the SAXS profiles of the as-synthesized and calcined DD samples. The as-synthesized sample exhibited several well-resolved reflections within the $q$ value range of 0.1-0.6 nm$^{-1}$. These reflections corresponded to the polymer matrix with a diamond minimal surface and the embedded DD TiO$_2$ networks with the space group $Pn\bar{3}m$ and the unit cell parameter of $a \approx 65$ nm. After calcination, the polymer template was removed, and the amorphous TiO$_2$ was transformed into crystalline anatase. However, the interpenetrating DD frameworks lost their mutual support, so the frameworks were unable to maintain their original cubic symmetry and thus adhered to each other along one of the <001> axes, therefore forming a tetragonal shifted DD skeleton. Therefore, the reflections with a $q$ ratio of √3:√8:√11 were indexed to the 101, 200/112, 121/103 reflections of the tetragonal shifted DD structure with unit cell parameters of $a = b = 70$ nm and $c = 99$ nm and a $c/a$ ratio of ~√2.

The SEM images of the as-synthesized sample showed a single crystal domain with a highly ordered polymer matrix with diamond surface structure (Fig. S13b and S13c). The high-resolution image indicates the typical {110} plane of the diamond minimal surface.

The SEM images of the calcined DD sample are shown in Fig. S13e and S13f. The high-resolution SEM image reveals two sets of interpenetrating frameworks, as indicated by the red and blue arrows. Each framework has tetragonal arrangements at the nodal sites and can be identified as diamond networks. Notably, these two sets of diamond networks closely adhered to each other due to the loss of mutual support.

The TEM images of the calcined DD sample taken from [110], [111] and [100] directions are presented in Fig. S14. The adhered double diamond networks can be revealed from the TEM images and the structural model. The TEM images simulations[4] and the stick models from different axes are shown as insets in the experimental TEM images and they were highly consistent with the observed TEM contrast.

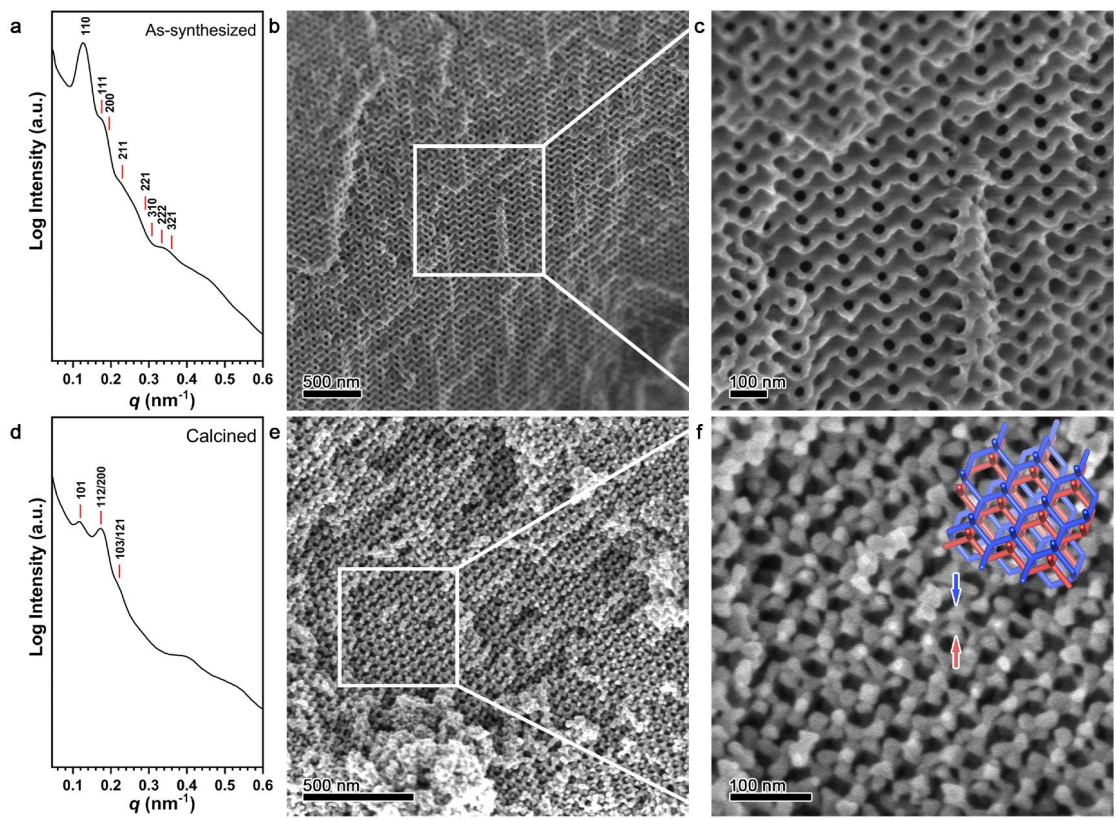

**Fig. S13. SAXS patterns and SEM images of the DD structure. a**, SAXS pattern of the as-synthesized DD sample. **b**, SEM image of the as-synthesized DD sample. **c**, High-magnification image of the selected region in **b** (white box). **d**, SAXS pattern of the calcined DD sample. **e**, SEM image of the calcined DD sample. **f**, High-magnification image of the selection shown in **e** (white box). The synthetic mixture included 0.05 g of PEO$_{45}$-*b*-PS$_{241}$, 8.0 g of THF, 1.2 g of HCl (3 M) and 0.5 g of TIA.

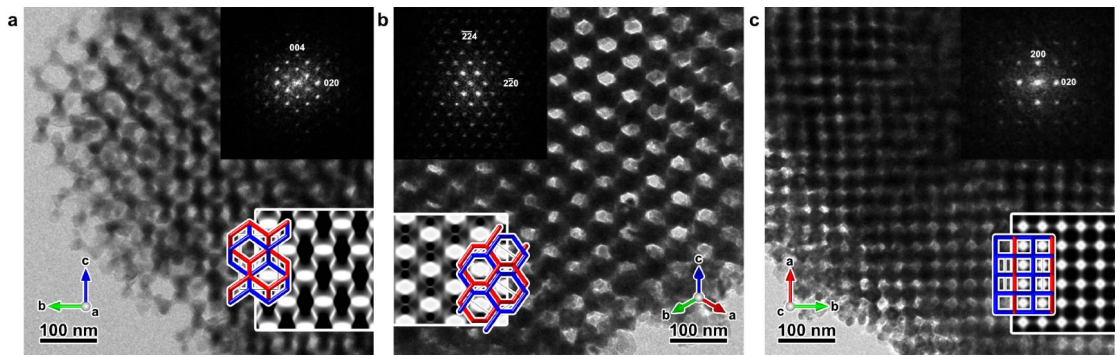

**Fig. S14. TEM images of the calcined DD structure. a-c**, TEM images and the corresponding FDs of the sample taken from the [110], [111] and [100] directions, respectively. During the calcination process, the interpenetrating DD frameworks lost the mutual support and the frameworks were shifted and adhered to each other, forming a low symmetry tetragonal lattice. The insets show TEM images simulated with a 3-term nodal equation. The schematic drawings of the shifted DD frameworks are overlaid on the TEM images and the simulation. The synthetic mixture included 0.05 g of $PEO_{45}$-*b*-$PS_{241}$, 8.0 g of THF, 1.2 g of HCl (3 M) and 0.5 g of TIA.

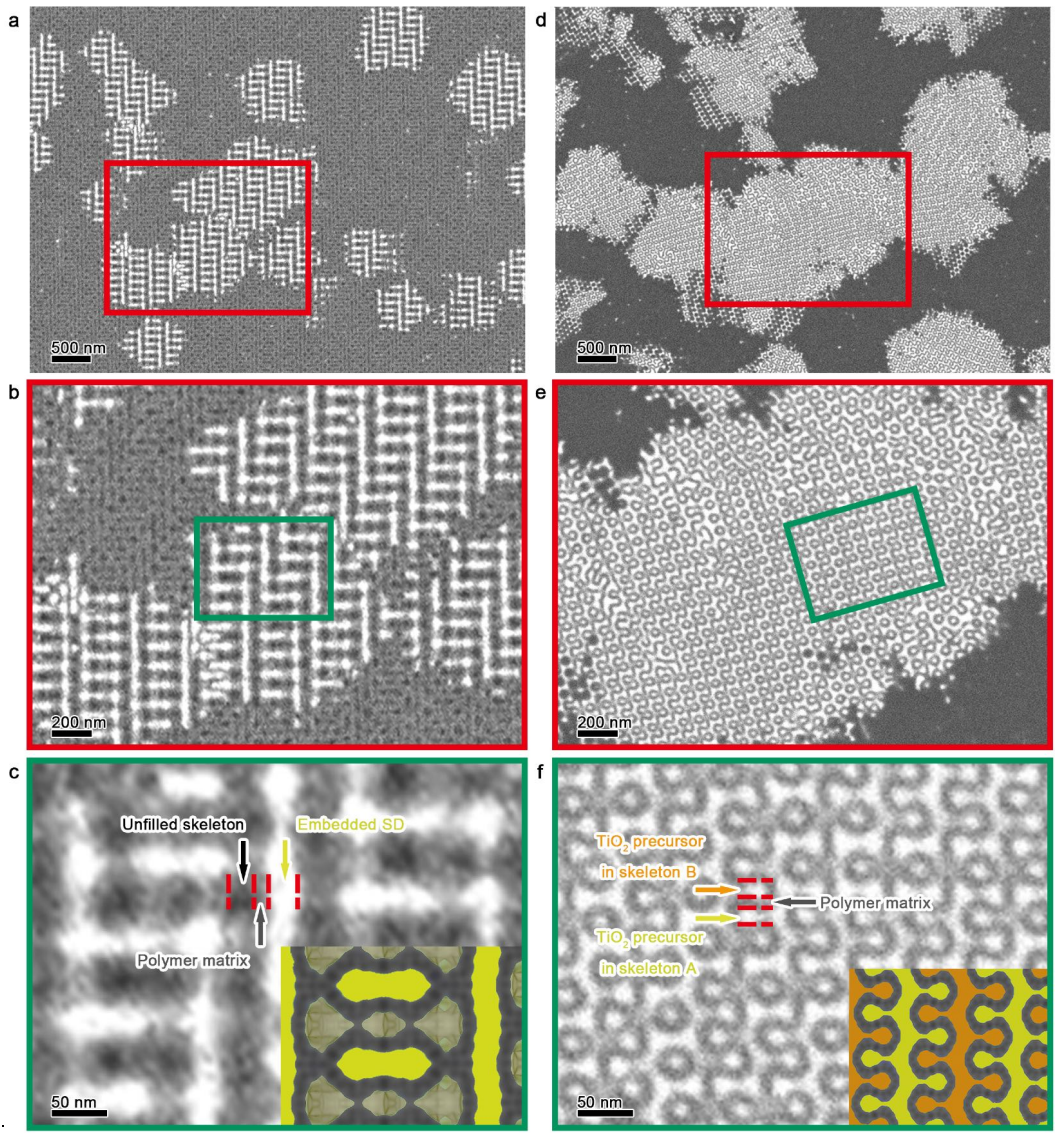

**Fig. S15. Cross-sectional BSE images of the as-synthesized SD and DD samples. a-c**, as-synthesized SD sample. **d-f**, as-synthesized DD sample. The synthetic mixture for SD included 0.05 g of PEO$_{45}$-*b*-PS$_{241}$, 8.0 g of THF, 0.8 g of HCl (3 M) and 0.5 g of TIA. The synthetic mixture for DD included 0.05 g of PEO$_{45}$-*b*-PS$_{241}$, 8.0 g of THF, 1.2 g of HCl (3 M) and 0.5 g of TIA.

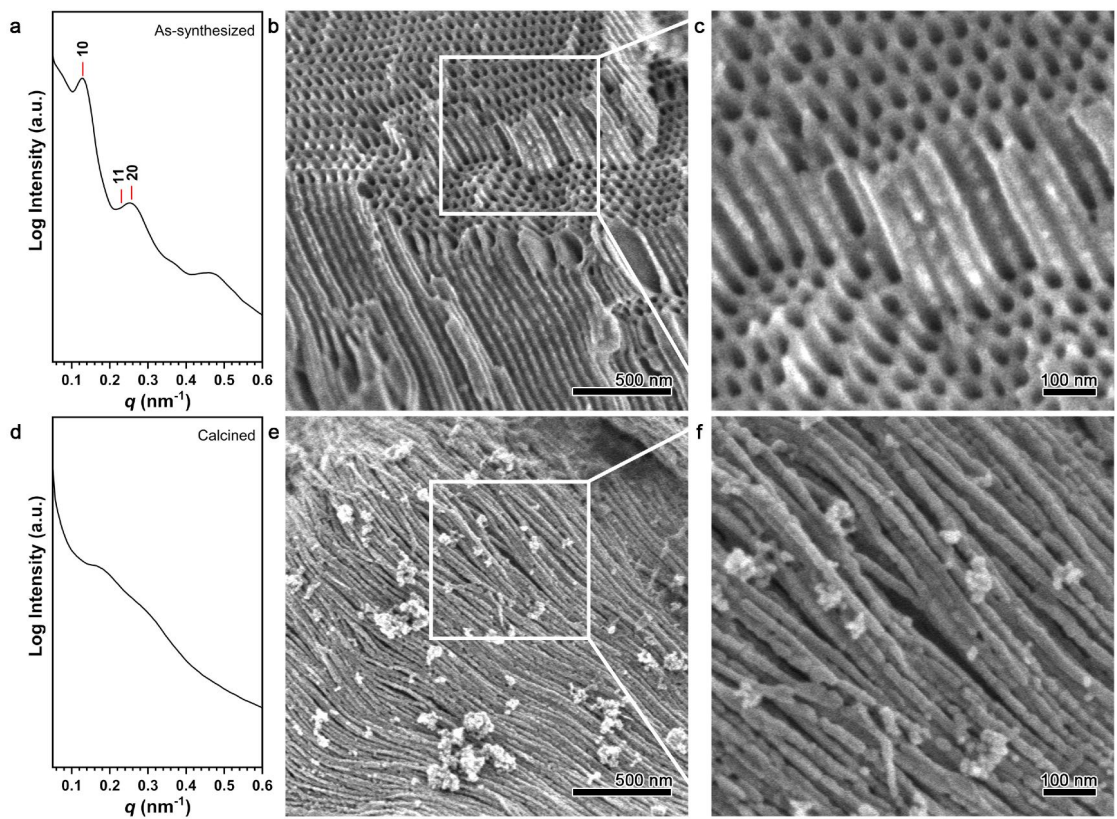

**Fig. S16. SAXS patterns and SEM images of the cylindrical structure. a**, SAXS pattern of the as-synthesized cylindrical structure. **b**, SEM image of the as-synthesized sample. **c**, High-magnification image of the corresponding area shown in **b** (white box). **d**, SAXS pattern of the calcined cylindrical structure. **e**, SEM image of the calcined material. **f**, High-magnification image of the corresponding area shown in **e** (white box). The synthetic mixture included 0.05 g of PEO$_{45}$-*b*-PS$_{241}$, 8.0 g of THF, 2.0 g of HCl (3 M) and 0.5 g of TIA.

Fig. S17 shows the SEM images of the freeze-dried samples and these were determined as a function of volatilization time. A small amount of solid product with a disordered network structure first appeared at 84 h. After solvent evaporation, the disordered structure was gradually transformed into diamond surface, and the order of the structure was improved after 100 h. Pure diamond surface structures were obtained after 112 h, and they showed the characteristic crystalline facets. Therefore, formation of the DD networks was explained as follows. The initial synthetic solution was uniform due to the large amount of the common solvent THF used. With volatilization of the solvent (mainly THF), the block copolymer gradually precipitated and aggregated to form the initial structure. With continued volatilization, the diamond surface matrix was formed by microphase separation of the block copolymer with the titania species embedded in the polymer matrix.

Fig. S18 shows the EDS maps of the samples synthesized in the time-course experiment shown in Fig. S17. The $TiO_2$ species were uniformly dispersed throughout the sample during the co-assembly process, suggesting that the $TiO_2$ precursor contributed to the initial volume fraction of the hydrophilic segment of the block copolymer to form the diamond minimal surface structure, which further condensed inside the networks to form the $TiO_2$ networks.

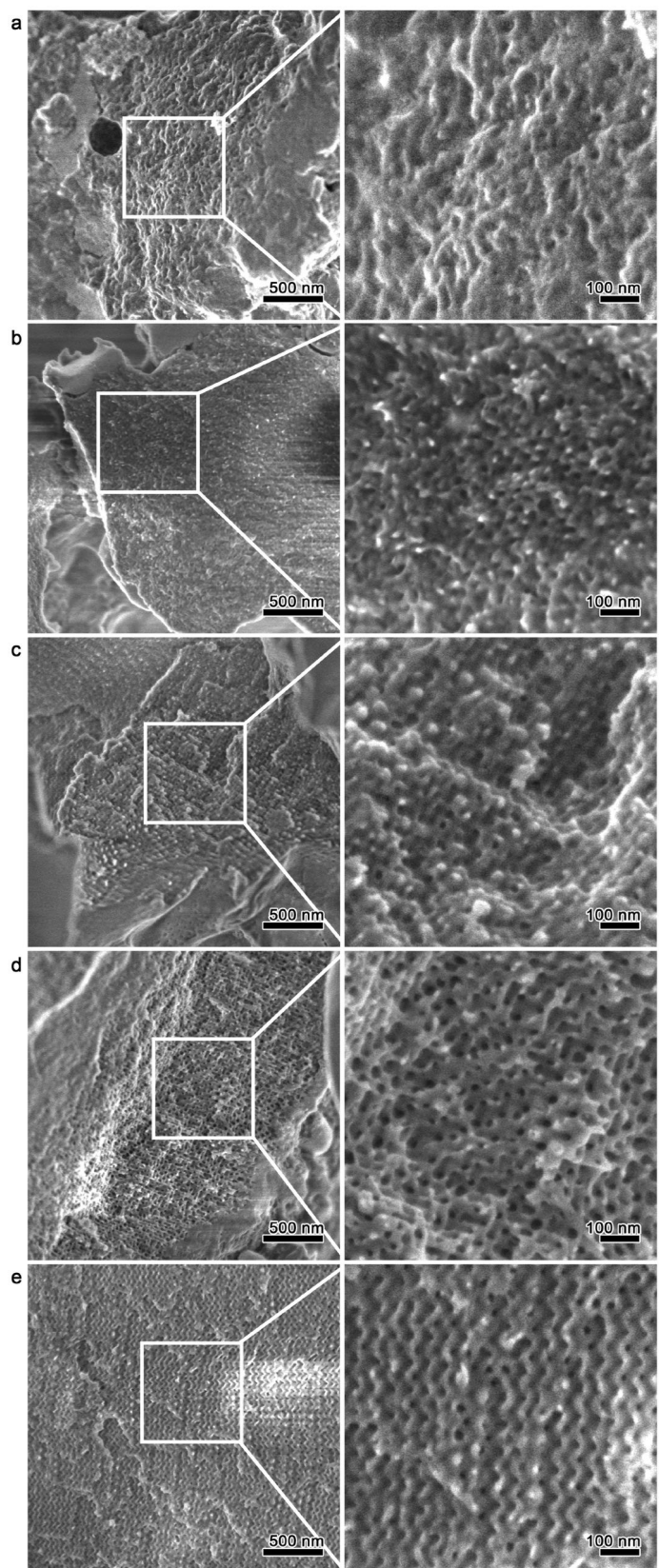

**Fig. S17. SEM images of the as-synthesized SD samples freeze-dried at different reaction times. a**, 84 h, **b**, 92 h, **c**, 100 h, **d**, 112 h and **e**, final product. The synthetic mixture included 0.05 g of PEO$_{45}$-*b*-PS$_{241}$, 8.0 g of THF, 0.8 g of HCl (3 M) and 0.5 g of TIA.

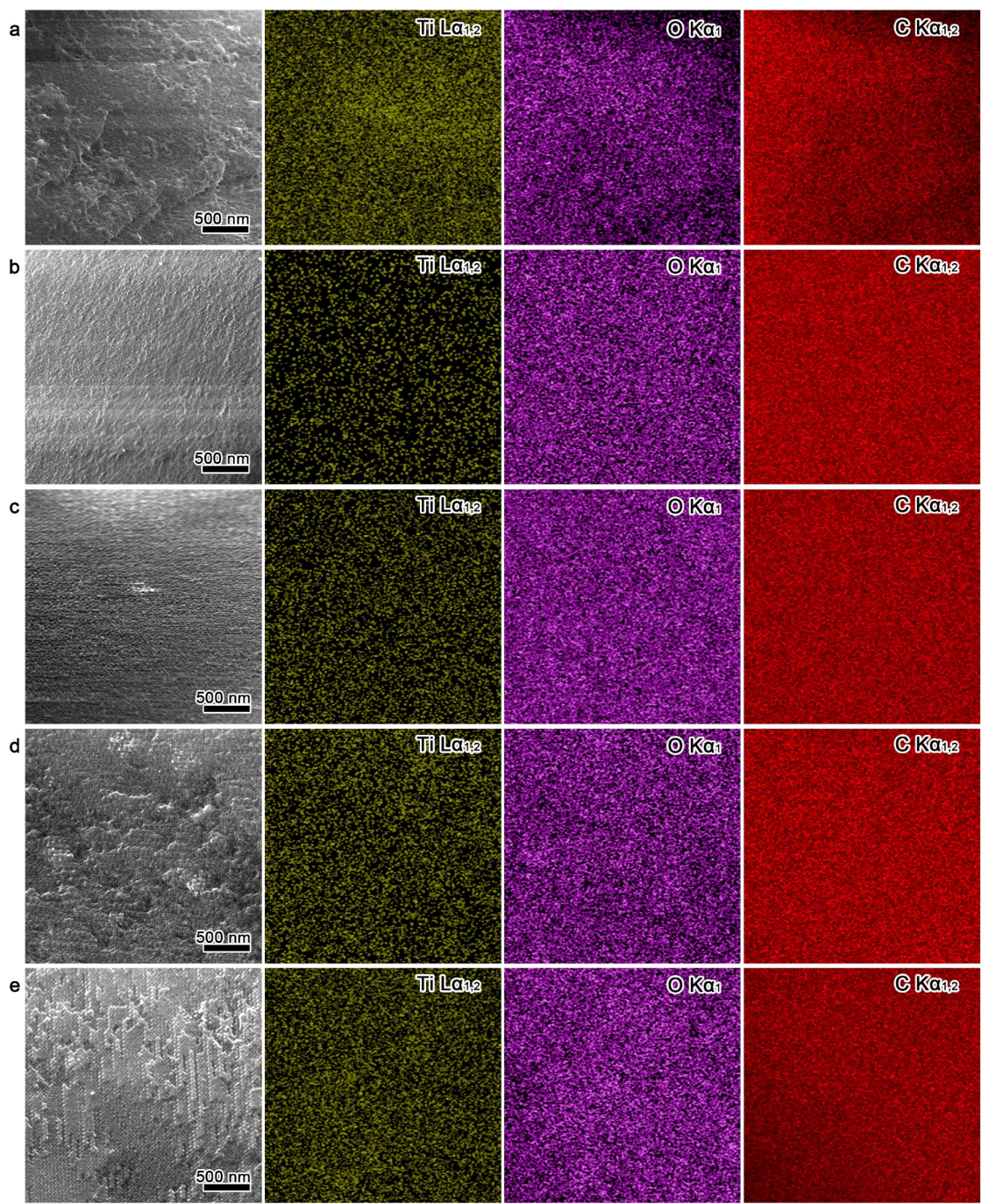

**Fig. S18. EDS maps of the as-synthesized SD samples freeze-dried at different reaction times. a**, 84 h, **b**, 92 h, **c**, 100 h, **d**, 112 h and **e**, final product. The synthetic mixture included 0.05 g of $PEO_{45}$-*b*-$PS_{241}$, 8.0 g of THF, 0.8 g of HCl (3 M) and 0.5 g of TIA.